\documentclass{article}

\usepackage[numbers]{natbib}
\usepackage{a4}
\usepackage{times}
\usepackage{amsmath}
\usepackage{amssymb}
\usepackage{latexsym}

\def \D{{\mathrm{D}}}

\def \pd{\partial}

\def \tl#1{\overset{\kern 1pt\circ}{#1}}
\def \TL#1{\overset{\kern -3pt \circ}{#1}}
\def \TLL#1{\overset{\kern -7pt \circ}{#1}}

\def \Bphi{\boldsymbol{\phi}}

\def \Bu{{\boldsymbol{u}}}

\def \Bx{{\boldsymbol{x}}}

\title{\bf BALANCE LAWS IN MICROMORPHIC ELASTICITY}

\author{ {\sl Markus Lazar, Charalampos Anastassiadis} \\[5pt]
        Emmy Noether Research Group,\\
        Department of Physics,\\
        Darmstadt University of Technology,\\   
        Hochschulstr. 6,\\      
        D-64289 Darmstadt, Germany\\[5pt]
        lazar@fkp.tu-darmstadt.de
}

\begin{document}

\maketitle

\paragraph {ABSTRACT:} We derive the Eshelby stress tensor, the angular momentum tensor and the dilatational vector flux for micromorphic elasticity. 
We give the corresponding balance laws and the $J$, $L$, and $M$ integrals.
Also we discuss when the balance laws become conservation laws.


\section{INTRODUCTION}
Conservation and balance laws are a very important research field in
theories of elasticity, generalized elasticity, material science
and engineering science~\cite{Maugin}.
Each infinitesimal symmetry group of the strain energy 
is associated with a conservation law~\cite{Olver}. 

In this paper we consider the microcontinuum field theory of 
micromorphic elasticity.
A micromorphic theory can be used to model materials with 
microstructure.
A micromorphic continuum, as we consider, is built up from
particles which possess an inherent orientation.
It possesses twelve degrees of freedom:
three translational ones and nine ones of the microstructure (rotation,
shear, dilatation). 

It is the purpose of the present paper to derive balance and conservation laws
for micromorphic elasticity. We take account for material inhomogeneity, anisotropy, body forces and body couples.
Also we will discuss the corresponding $J$, $L$ and $M$ integrals.

\section{MICROMORPHIC ELASTICITY}
We consider the theory of micromorphic elasticity~\citep{Mindlin64,ES64,Eringen99}.
Micromorphic elasticity views a material as a continuous collection of deformable
particles. Each particle is attached with a microstructure of finite size. 
In addition to macro-deformations,
we have to consider 
micro-deformations
in order to describe microstructural effects
(micro-rotation, micro-dilatation and micro-shear of the microstructure).
The basic macro-field is the displacement vector $\Bu(\Bx)$ 
and the basic micro-field is the so-called micro-distortion tensor $\Bphi(\Bx)$.
Let the strain energy density of micromorphic 
elasticity be of the form\footnote{The usual notations
$A_{,i}=\pd A/(\pd x_i)$ and  $\dot A=\pd A/(\pd t)$ are used.}
\begin{align}
\label{align}
W=W(x_i,u_\alpha,\phi_{\alpha\beta}, u_{\alpha,i},\phi_{\alpha\beta,i}).
\end{align}
The Euler-Lagrange equations are obtained according to
\begin{align}
\label{E-1}
E_\alpha(W)&=\frac{\pd W}{\pd u_\alpha}
-\D_i\, \frac{\pd W}{\pd u_{\alpha,i}}=0,\\
\label{E-2}
E_{\alpha\beta}(W)&=\frac{\pd W}{\pd \phi_{\alpha\beta}}
-\D_i\, \frac{\pd W}{\pd \phi_{\alpha\beta,i}}=0,
\end{align}
where $\D_i$ is the so-called total derivative~\cite{Olver} 
\begin{align}
\D_i=\frac{\pd}{\pd x_i}
+u_{\alpha,i}\,\frac{\pd}{\pd u_\alpha}
+u_{\alpha,ij}\,\frac{\pd}{\pd u_{\alpha,j}}
+\phi_{\alpha\beta,i}\,\frac{\pd}{\pd \phi_{\alpha\beta}}
+\phi_{\alpha\beta,ij}\,\frac{\pd}{\pd \phi_{\alpha\beta,j}}
+\dots\ .
\end{align}

In linear micromorphic elasticity, the strains are related to a
displacement vector and a micro-distortion tensor according 
\begin{align}
\label{def}
\gamma_{kl}=u_{k,l}-\phi_{kl},\qquad
2e_{kl}=\phi_{kl}+\phi_{lk},\qquad
\kappa_{klm}=\phi_{kl,m}.
\end{align}
Here $\gamma_{kl}$, $e_{kl}$ and $\kappa_{kl}$ are the relative
distortion, the micro-strain and the wryness tensors.
Using the strain tensors,
the strain energy density with external sources is of the form
\begin{align}
\label{W2}
W=\frac{1}{2}\, t_{kl}\gamma_{kl}+\frac{1}{2}\,s_{kl} e_{kl}+\frac{1}{2}\, m_{ijk}\kappa_{ijk} -u_\alpha F_\alpha -\phi_{\alpha\beta} L_{\alpha\beta}.
\end{align}
For linear and anisotropic micromorphic elasticity, 
the constitutive relations for the stresses are 
\begin{align}
\label{CE1}
t_{ij}&=\frac{\pd W}{\pd \gamma_{ij}}
=A_{ijkl}\gamma_{kl}+E_{ijkl} e_{kl}+F_{ijklm} \kappa_{klm} ,\\
\label{CE2}
s_{ij}&=\frac{\pd W}{\pd e_{ij}}
=E_{klij}\gamma_{kl}+B_{ijkl} e_{kl}+G_{ijklm} \kappa_{klm} ,\\
\label{CE3}
m_{ijk}&= \frac{\pd W}{\pd \kappa_{ijk}}
=F_{lmijk}\gamma_{lm}+G_{lmijk} e_{lm}
+C_{ijklmn}\kappa_{lmn}
\end{align}
with $s_{ij}=s_{ji}$.
Here $t_{ij}$ is the force stress tensor, 
$s_{kl}$ is the micro-stress tensor and $m_{klm}$ is the stress moment tensor.
The tensors $A_{ijkl}$, $B_{ijkl}$, $C_{ijklmn}$, $E_{ijkl}$, $F_{ijklm}$
and $G_{ijklm}$ are constitutive tensors.
They fulfill the symmetry relations:
\begin{align}
\label{sym}
&A_{ijkl}=A_{klij},\quad
B_{ijkl}=B_{klij}=B_{jikl}=B_{ijlk},\quad
C_{ijklmn}=C_{lmnijk},\nonumber\\
&E_{ijkl}=E_{jikl},\quad
G_{ijklm}=G_{jiklm}.
\end{align}

The external body forces $F_\alpha$ and body couples $L_{\alpha\beta}$ 
are defined by
\begin{align}
\label{F}
F_\alpha:=-\frac{\pd W}{\pd u_\alpha},\qquad
L_{\alpha\beta}:=-\frac{\pd W}{\pd \phi_{\alpha\beta}}.
\end{align}
Therefore, the forces and couples are conservative.
In addition, the Lagrangian may
depend explicitly  on $x_i$.
In this case, the material force (or inhomogeneity force) 
is defined by
\begin{align}
f^{\text{inh}}_i:=-\frac{\pd W}{\pd x_i},
\end{align}
which is caused by material inhomogeneities.

Using eqs.~(\ref{E-1}) and (\ref{E-2}), the Euler-Lagrange equations read
in terms of the canonical conjugate quantities:
\begin{align}
\label{E-1-2}
&\D_i t_{\alpha i}+F_\alpha=0 ,\\
\label{E-2-2}
&\D_i m_{\alpha\beta i}+(t_{\alpha\beta}-s_{\alpha\beta})+L_{\alpha\beta}=0 .
\end{align}


\section{TRANSLATIONAL, ROTATIONAL AND DILATATIONAL FLUXES}

In micromorphic elasticity the flux is given by~\cite{LA06}
\begin{align}
\label{A}
A_i=U_\alpha\, \frac{\pd W}{\pd u_{\alpha,i}}+\Phi_{\alpha\beta}\, \frac{\pd W}{\pd \phi_{\alpha\beta,i}}+X_i W
-X_j\bigg(u_{\alpha,j}\, \frac{\pd W}{\pd u_{\alpha,i}}+\phi_{\alpha\beta,j}\, \frac{\pd W}{\pd \phi_{\alpha\beta,i}}\bigg).
\end{align}
Here the infinitesimal generators are defined by
\begin{align}
X_i(\Bx,\Bu,\Bphi)&:=\frac{\pd x'_i}{\pd\varepsilon}\bigg|_{\varepsilon=0},\ 
U_\alpha(\Bx,\Bu,\Bphi)&:=\frac{\pd u'_\alpha}{\pd\varepsilon}\bigg|_{\varepsilon=0},
\
\Phi_{\alpha\beta}(\Bx,\Bu,\Bphi):=\frac{\pd \phi'_{\alpha\beta}}{\pd\varepsilon}\bigg|_{\varepsilon=0}
\end{align}
and the group action is of the form
\begin{align}
x'_i&=x_i+\varepsilon X_i(\Bx,\Bu,\Bphi)+\cdots\ ,\\
u'_\alpha&=u_\alpha+\varepsilon U_\alpha(\Bx,\Bu,\Bphi)+\cdots\ ,\\
\phi'_{\alpha\beta}&=\phi_{\alpha\beta}+\varepsilon \Phi_{\alpha\beta}(\Bx,\Bu,\Bphi)+\cdots\ ,
\end{align}
where $\varepsilon$ is a group parameter.

\subsection{Translations}
The translation in space is given by the formulae:
\begin{align}
x'_i =x_i+\varepsilon_k \delta_{ki},\qquad
u'_\alpha=u_\alpha,\qquad
\phi'_{\alpha\beta}=\phi_{\alpha\beta} .
\end{align}
The corresponding generators of infinitesimal transformations are
\begin{align}
\label{gen-transl}
X_{ki}=\delta_{ki},\qquad U_\alpha=0,\qquad \Phi_{\alpha\beta}=0 .
\end{align}
Using eqs.~(\ref{A}) and (\ref{gen-transl}), 
the translational flux is given by
\begin{align}
\label{EMT0}
P_{ki}&=W\delta_{ki}
-u_{\alpha,k}\,\frac{\pd W}{\pd u_{\alpha,i}}-\phi_{\alpha\beta,k}\, \frac{\pd W}{\pd \phi_{\alpha\beta,i}}.
\end{align}
In terms of stresses it reads
\begin{align}
\label{EMT}
P_{ki}&=W\delta_{ki}
-u_{\alpha,k}\, t_{\alpha i}-\phi_{\alpha\beta,k}\, m_{\alpha\beta i} .
\end{align}
Eq.~(\ref{EMT}) is nothing but the Eshelby stress tensor for micromorphic elasticity~\cite{LA06,Lazar07,LM07}.

\subsection{Rotations}
The three-dimensional group of rotations acts in the space of
the independent $\Bx$ and dependent variables $\Bu$ and $\Bphi$. 
Its infinitesimal action is given by:
\begin{align}
x'_i=x_i+\epsilon_{kji} x_j \varepsilon_k,\
u'_\alpha=u_\alpha+\epsilon_{k \beta \alpha} u_\beta \varepsilon_k,\ 
\phi'_{\alpha\beta}=\phi_{\alpha\beta}
+\epsilon_{kj \alpha } \phi_{j\beta} \varepsilon_k
+\epsilon_{kj \beta } \phi_{\alpha j} \varepsilon_k .
\end{align}
Eventually, the infinitesimal generators are easily obtained  
\begin{align}
\label{gen-rot}
X_{ik}=\epsilon_{ikj}x_j,\qquad 
U_{\alpha k}=\epsilon_{\alpha k\beta}u_\beta,\qquad
\Phi_{\alpha\beta k}=\epsilon_{\alpha kj } \phi_{j\beta}+\epsilon_{\beta kj } \phi_{\alpha j} .
\end{align}
If we use eqs.~(\ref{A}) and (\ref{gen-rot}), 
the rotational flux is given by
\begin{align}
\label{AMT0}
M_{ki}=\epsilon_{kji}x_j W
&+\epsilon_{kj\alpha}\bigg(u_j\, \frac{\pd W}{\pd u_{\alpha,i}}
+\phi_{jl}\, \frac{\pd W}{\pd\phi_{\alpha l,i}}
+\phi_{lj}\, \frac{\pd W}{\pd\phi_{l\alpha ,i}}
\bigg)\nonumber\\
&-\epsilon_{kjn} x_j\bigg(u_{\alpha,n}\, \frac{\pd W}{\pd u_{\alpha,i}}
+\phi_{\alpha\beta,n}\, \frac{\pd W}{\pd\phi_{\alpha\beta,i}}\bigg) .
\end{align}
In terms of the Eshelby tensor~(\ref{EMT}) and the stress tensors we obtain
\begin{align}
\label{AMT}
M_{ki}&=\epsilon_{kjn}\big(x_j P_{ni}+u_j\, t_{ni}
+\phi_{lj}\, m_{lni}+\phi_{jl}\, m_{nli}\big).
\end{align}
It is the total angular momentum tensor.
It can be decomposed into the orbital and intrinsic (spin) angular momentum 
tensors
\begin{align}
M_{ki}=M^{\text{(o)}}_{ki}+M^{\text{(i)}}_{ki} .
\end{align}
The orbital angular momentum tensor reads
\begin{align}
M^{\text{(o)}}_{ki}=\epsilon_{kjn}\, x_j P_{ni} .
\end{align}
The intrinsic (spin) angular momentum tensor is given by
\begin{align}
M^{\text{(i)}}_{ki}=\epsilon_{kjn}\big( u_j t_{n i}
+\phi_{lj} m_{l n i}+\phi_{jl} m_{nl  i}\big).
\end{align}

\subsection{Scaling}

The scaling group acts in infinitesimal form 
on the independent and dependent variables
\begin{align}
x'_i=(1+\varepsilon)  x_i,\quad
u'_\alpha&=(1+\varepsilon\, d_u) u_\alpha,\quad
\phi'_{\alpha\beta}=(1+\varepsilon\, d_\phi) \phi_{\alpha\beta} ,
\end{align}
where $d_u$ and $d_\phi$ denote the (scaling) dimensions 
of the displacement vector $\Bu$ and the micro-deformation tensor $\Bphi$.
The infinitesimal generators are given by
\begin{align}
\label{Gen-s}
X_i=x_i,\quad U_\alpha=d_u u_\alpha,\quad
\Phi_{\alpha\beta}= d_\phi\phi_{\alpha\beta} ,
\end{align}
where
\begin{align}
\label{SD}
d_u=-\frac{n-2}{2},\qquad
d_\phi=-\frac{n}{2},
\end{align}
and $\delta_{ll}=n$.
If we substitute (\ref{Gen-s}) into (\ref{A}),
we obtain for the scaling flux
\begin{align}
\label{SV-0}
Y_{i}&=x_i W
+\big(d_u u_{\alpha} -x_k u_{\alpha,k}\big)\,
\frac{\pd W}{\pd u_{\alpha,i}}
+\big(d_\phi \phi_{\alpha\beta} -x_k \phi_{\alpha\beta,k}\big)\,
\frac{\pd W}{\pd \phi_{\alpha\beta,i}}.
\end{align}
In terms of the stress tensors the scaling flux reads
\begin{align}
\label{Y}
Y_{i}=x_j P_{ji}+ d_u u_j\, t_{j i}
+d_\phi \phi_{jl}\, m_{jli}.
\end{align}


\section{BALANCE AND CONSERVATION LAWS}
We now turn to the discussion of the properties of the fluxes.
The information of a symmetry defined by the transformation law
of the fields lies in the properties of the divergence of the corresponding
currents. If the divergence is zero, we speak of a conservation law.
If it is not zero, we have a balance law.

For  micromorphic elasticity we obtain 
with the corresponding Eshelby stress tensor~(\ref{EMT}), total angular momentum tensor~(\ref{AMT})
and scaling flux vector~(\ref{Y}) the following balance laws:
\begin{align}
\label{BL-P}
&\D_i P_{ki}=-f^{\text{inh}}_k ,\\
\label{BL-rot}
&\D_i M_{ki}=\epsilon_{kjn}\big(
\gamma_{ij} t_{in}+\gamma_{ji}\, t_{ni}
+2e _{ij}s_{in}
+\kappa_{ijl}\, m_{inl}
+\kappa_{jli}\, m_{nli}
+\kappa_{lij}\, m_{lin}
\big)\nonumber\\
&\qquad\qquad
-\epsilon_{kjn}\big( x_j f^{\text{inh}}_n
+u_j F_n +\phi_{ji}L_{ni}+\phi_{ij}L_{in}\big),\\
\label{BL-Y}
&\D_i Y_i=-\kappa_{\alpha\beta i}\, m_{\alpha\beta i}
-x_i f^{\text{inh}}_i
-\frac{n+2}{2}\, u_\alpha F_\alpha -\frac{n}{2}\, \phi_{\alpha\beta} L_{\alpha\beta}.
\end{align}
Eq.~(\ref{BL-P}) follows 
essentially from the lack of translational invariance and it 
can be called the canonical momentum balance law.
In eq. (\ref{BL-rot}) the terms in the first parentheses on the right hand side vanish 
if the micromorphic material is isotropic~\cite{LA06}.
The other terms on the right hand side are resulting vector moments 
caused by inhomogeneities, external body forces and couples.
Eq.~(\ref{BL-rot}) may be called the canonical angular momentum balance law.
The source terms in eq.~(\ref{BL-Y})
are `scalar moments' breaking the scaling symmetry.
The fist source term breaks 
the dilatational symmetry due to the stress moment tensor $m_{ijk}$.
Additional source terms appear which account for material inhomogeneities, external forces and 
couples.
Eq.~(\ref{BL-Y}) can be called the scalar moment of momentum balance law.

In integral form we obtain the $J$, $L$ and $M$ integrals of
micromorphic elasticity:
\begin{align}
J_k&=\int_S P_{ki} n_i\, d S =-\int_V f^{\text{inh}}_k\, d V,\\
L_k&=\int_S M_{ki} n_i\, d S=\int_V \epsilon_{kjn}\big[
\gamma_{ij} t_{in}+\gamma_{ji}\, t_{ni}
+2e _{ij}s_{in}
+\kappa_{ijl}\, m_{inl}
+\kappa_{jli}\, m_{nli}
\nonumber\\
&\hspace{2.5cm}
+\kappa_{lij}\, m_{lin}
-\big( x_j f^{\text{inh}}_n
+u_j F_n +\phi_{ji}L_{ni}+\phi_{ij}L_{in}\big)\big]\, d V,\\
M&=\int_S Y_{i} n_i\, d S =
-\int_V\big(\kappa_{\alpha\beta i}\, m_{\alpha\beta i}
+x_i f^{\text{inh}}_i+\frac{n+2}{2}\, u_\alpha F_\alpha 
+\frac{n}{2}\, \phi_{\alpha\beta} L_{\alpha\beta}\big)\, d V.
\end{align}
For an isotropic material with vanishing
body forces, body moments and material forces 
we obtain:
\begin{align}
J_k&=\int_S P_{ki} n_i\, d S =0,\\
L_k&=\int_S M_{ki} n_i\, d S=0,\\
M&=\int_S Y_{i} n_i\, d S =
-\int_V \kappa_{\alpha\beta i}\, m_{\alpha\beta i} \, d V.
\end{align}
Thus, in this case the $J$ and $L$ integrals are zero and
the Eshelby stress and total angular momentum tensors 
give conservation laws because the are divergence-less.
On the other hand, the $M$ integral is non-zero
because the term $m_{ijk}$ breaking the scaling symmetry survives.
Therefore, it is not a conservation integral.

\section*{Acknowledgement}
The authors have been supported by an Emmy-Noether grant of the 
Deutsche Forschungsgemeinschaft (Grant No. La1974/1-2).

\bibliographystyle{unsrt}

\end{document}